# Plasmon Lifetime in K: A Case Study of Correlated Electrons in Solids Amenable to *Ab Initio* Theory

*Wei Ku and Adolfo G. Eguiluz*

Department of Physics and Astronomy, The University of Tennessee, Knoxville, TN 37996-1200
and Solid State Division, Oak Ridge National Laboratory, Oak Ridge, TN 37831-6030

**Abstract**

On the basis of a new *ab initio*, all-electron response scheme, formulated within time-dependent density-functional theory, we solve the puzzle posed by the anomalous dispersion of the plasmon linewidth in K. The key damping mechanism is shown to be decay into particle-hole pairs involving empty states of *d*-symmetry. While the effect of many-particle correlations is small, the correlations built into the "final-state" *d*-bands play an important, and novel, role —which is related to the phase-space complexity associated with these flat bands. Our case study of plasmon lifetime in K illustrates the importance of *ab initio* paradigms for the study of excitations in correlated-electron systems.

Much less is known about how to calculate accurate lifetimes of electronic excitations for realistic models of solids than about ground-state and structural properties of these systems. Currently a major thrust is focused on the excitations (involving spin, charge, phononic degrees of freedom, etc.) in systems of strongly-correlated electrons. The complex band structure characteristic of these systems notwithstanding, the prevalent paradigm adopted for their study invokes sophisticated treatments of correlation together with minimal representations of the band structure [1]. Interestingly, for systems with simpler band structures the puzzles posed by a variety of spectroscopic data [2,3] have led to the advocacy of a similar paradigm. This being the case, the latter systems offer the possibility —which we pursue in this work— that, on account of their relative simplicity, they may be more amenable to *ab initio* theory.

As a significant example we have that landmark electron energy-loss spectroscopy (EELS) measurements by vom Felde, Sprösser-Prou, and Fink [2,4] revealed remarkable anomalies in the excitation spectrum of the alkali metals. For example [5], the dispersion of the plasmon linewidth in K (and to a lesser degree in Na) was found to differ qualitatively from theoretical predictions obtained on the premise that the effects of the one-electron band structure are unimportant, except for the presence of a small gap just above the Fermi surface [6]. On this basis it was concluded that *(i)* the linewidth dispersion is anomalous, and *(ii)* the physics behind the anomaly must be due to strong dynamical short-range correlations [2]. Since these correlations vanish in the high-density limit $r_S \to 0$, which is far from being realized in K, in which the density is low ($r_S \sim 5$), conclusion *(ii)* seems "natural," and is consistent with the absence of single-particle fine structure in the EELS data.

In this Letter we show that the anomalous dispersion of the plasmon linewidth in K is, in fact, controlled by decay into single particle-hole pairs involving empty states of *d*-symmetry. By comparison with the EELS data we conclude that dynamical many-body correlations (which we turn off), do not affect the linewidth dispersion. By contrast, the exchange-correlation effects built into the crucial "final-state" *d*-bands have a profound impact on the calculated linewidths. The identification of the physics behind the data is made possible by our approaching the problem on the basis of a new all-electron response scheme which is framed within time-dependent density-functional theory (TDDFT) [7,8]. The response of Kohn-Sham electrons is evaluated for imaginary frequencies, followed by analytic continuation to the real axis, which we perform numerically; this procedure leads us to an accurate extraction of the natural linewidth. The novel features of our results are related to the phase space complexity introduced by the flat *d*-bands which dominate the physics of the damping; this complexity is absent in popular models of correlated-electron systems [1].

On first glance, it may seem that the energy scale of the physics involved in plasmon damping (~ 4 eV, in K) is too large to be of significance for the current emphasis in the study of low-energy excitations and correlations near the Fermi surface. However, in fully *self-consistent* many-body calculations [9] the collective mode feeds back onto the one-particle spectrum, and this effect influences, e.g., the value of the quasiparticle renormalization factor at the Fermi surface. Moreover, the electronic correlations at low energies are of course affected by screening, and this process —usually bypassed via phenomenological on-site interactions— is closely related to the physics of the collective mode.

In TDDFT the *exact* linear density-response function $\chi(\vec{x},\vec{x}';\omega)$ obeys the integral equation [7,8]

$$\chi = \chi^{(s)} + \chi^{(s)}(v + f_{xc})\chi \ , \tag{1}$$

where $\chi^{(s)}(\vec{x},\vec{x}';\omega)$ is the response function for "non-interacting" Kohn–Sham electrons, $v(\vec{x}-\vec{x}')$ is the Coulomb interaction, and $f_{xc}(\vec{x},\vec{x}';\omega)$ accounts for dynamical exchange-correlation effects. For a periodic crystal the Fourier transform of $\chi^{(s)}$ is given by

$$\chi^{(s)}_{\vec{G},\vec{G}'}(\vec{q};\omega) = \frac{1}{V}\sum_{\vec{k}}^{BZ}\sum_{n,n'}\frac{f_{\vec{k},n} - f_{\vec{k}+\vec{q},n'}}{E_{\vec{k},n} - E_{\vec{k}+\vec{q},n'} + \hbar(\omega + i\eta)}\left\langle \vec{k},n\right| e^{-i(\vec{q}+\vec{G})\cdot\hat{\vec{x}}}\left|\vec{k}+\vec{q},n'\right\rangle$$
$$\text{x} \quad \left\langle \vec{k}+\vec{q},n'\right| e^{i(\vec{q}+\vec{G}')\cdot\hat{\vec{x}}}\left|\vec{k},n\right\rangle \ , \tag{2}$$

where $\vec{G}$ is a vector of the reciprocal lattice, $n$ is a band index, the wave vectors $\vec{k}$ and $\vec{q}$ are in the first Brillouin zone (BZ), and $V$ is the normalization volume. In this representation Eq. (1) is turned into a matrix equation which we solve numerically; the plasmon loss corresponds to a well-defined peak in Im $\chi_{\vec{G}=0,\vec{G}'=0}(\vec{q};\omega)$ for a given $\vec{q}$. The off-diagonal matrix elements in Eq. (2) incorporate the "crystal-local fields" in the response function. This effect turns out to play a minor role in the present calculations; thus, it will not be mentioned any further.

In this work we compute $\chi^{(s)}$ making use of the local-density approximation (LDA) in the evaluation of the exchange-correlation potential $V_{xc}(\vec{x})$ [10]. The remarkable success of this approximation in the present response context is addressed below. We adopt an all-electron approach, which starts from the knowledge of the LDA band structure and wave functions [11] in the full-potential linearized augmented plane wave (LAPW) method [12].

We aim at disentangling the effects of the one-particle band structure from those of dynamical many-body correlations. The TDDFT linear-response framework [8] is well suited for this purpose, as it allows us to turn off the latter by setting $f_{xc} = 0$ —which we do [13]. Hence, we concentrate on elucidating the impact of single-particle decay channels.

Now, an accurate evaluation of the linewidth of a long-lived excitation such as the K plasmon poses a non-trivial problem to theory. Indeed, when sampling the BZ in Eq. (2), conventional methods introduce numerical broadening via a finite value of $\eta$ which typically is in the 0.5 eV range —larger than the natural linewidth. We proceed differently. We first compute $\chi^{(s)}$, and solve Eq. (1), for imaginary frequencies —this allows us to sample the BZ without numerical broadening; $\eta \equiv 0$ in Eq. (2). The physical response $\chi$ is obtained on the real axis —a distance $\delta$ above it— by analytic continuation, which we perform via Padé approximants [14]. The power of our method is illustrated in Fig. 1, in which we compare the loss function $\mathrm{Im}\,\chi_{\vec{G}=0,\vec{G}'=0}(\vec{q};\omega)$ obtained as just outlined [15] with its counterpart obtained via the (more standard) direct evaluation of Eq. (2) for real $\omega$'s (a finite $\eta$ is now required) [16]. Figure 1 corresponds to a 16x16x16 wave vector mesh [17]; for the same $\vec{k}$-mesh, the real-axis approach cannot resolve the linewidth. It is crucial that our calculated loss peak is quite insensitive to a decrease of $\delta$ by an order of magnitude. Since in the upper panel $2\delta$ is ~ 1/100 of the plasmon linewidth, an accurate value of the natural width is thus extracted [18].

In Fig. 2 we show a well-converged calculation of the plasmon linewidth dispersion of K (solid circles), obtained using a 20x20x20 $\vec{k}$-mesh and an energy cutoff of 20 eV in Eq. (2), corresponding to the inclusion of ~ 20 valence bands plus core states (the latter states are discussed towards the end of our presentation). Our result is compared with the EELS data of vom Felde et al. [2] (empty diamonds). Note that the full-width at half-maximum of the plasmon peak, $\Delta E_{1/2}(\vec{q})$, is given relative to its extrapolated value for $\vec{q} = 0$, a convention also adopted by previous authors [2,6]. Clearly, our results are in excellent agreement with experiment; since this agreement is obtained for $f_{xc} = 0$, we conclude that the plasmon linewidth *dispersion* of K is *not* controlled by a dynamical many-body mechanism [19,20].

The result of Fig. 2 is striking, as "intuitive" expectations based on the fact that the gap just above the Fermi surface at the *N*-point is small [21] yield the result shown by the solid line in Fig. 2, which corresponds to an evaluation of the dielectric function to second order in an empirical pseudopotential [6]. It is apparent that the use of nearly-free-electron states and eigenvalues in the evaluation of $\chi^{(s)}$ breaks down completely. Note that the fact that this simple-model result is so far off the EELS data is precisely what led to the proposal of the importance of dynamical correlations [2].

In Fig. 3 we analyze the role of key "final state" bands entering the evaluation of the single-particle response $\chi^{(s)}$; the relevant part of the band structure is shown in the inset of Fig. 3 (left panel), in which the shaded strip is the $\omega$-interval representing all the single-particle states which may couple to the plasmon (as determined by the conservation laws of energy and crystal momentum) [22]. Now *three* valence bands (thin solid lines) are needed in order to obtain a good dispersion curve for the plasmon *energy* [23]. Keeping just these bands, our plasmon linewidth dispersion curve (solid triangles) agrees well with the result of Ref. [6], which is understandable, as the states kept are, for the most part, nearly-free-electron-like.

Next, we have the result (solid squares) which incorporates the contribution from three additional bands (thick solid lines in the inset). The inclusion of these bands brings about a *qualitative* change in the plasmon linewidth dispersion curve, which is now quite close to the EELS data [24]. *Clearly, these three bands provide the key decay channels for the plasmon of K.* It is significant that the bands in question are overwhelmingly of *d*-character, as evidenced by the angular momentum-resolved density of states (DOS) shown on the right panel of Fig. 3.

The damping channels we have just identified yield a *positive* plasmon linewidth dispersion. We can visualize this point with the aid of the qualitative result that [6]

$$\Delta E_{1/2}(\vec{q}) \approx \operatorname{Im}\varepsilon(\vec{q};\omega)\left(\partial \operatorname{Re}\varepsilon(\vec{q};\omega)/\partial\omega\right)^{-1}\Big|_{\omega=\omega_p(\vec{q})} , \quad (3)$$

where $\varepsilon(\vec{q};\omega) = 1 - v(q)\chi^{(s)}(\vec{q};\omega)$ is the dielectric function (its $\vec{G}=0, \vec{G}'=0$ element). The first factor in Eq. (3) represents the availability of channels for decay into particle-hole pairs with energy $\hbar\omega_p(\vec{q})$ (given by the position of the peak in $\operatorname{Im}\chi(\vec{q};\omega)$); the second represents a "polarization effect" from interband transitions taking place at other $\omega$'s. Then, since $\omega_p(\vec{q})$ disperses upwards [23], a larger portion of the flat bands in the shaded strip becomes available to the plasmon as $|\vec{q}|$ increases, which naturally leads to an enhancement of the first factor in Eq. (3) —thus the positive linewidth dispersion, in agreement with experiment.

We concluded above that dynamical many-body correlations are unimportant in the present problem. However, there *is* a subtle many-body effect built into our result of Fig. 2; the same has to do with the exchange-correlation effects included in the Kohn-Sham response $\chi^{(s)}$. We emphasize that, although the single-particle states entering Eq. (2) —in particular, the crucial *d*-bands of interest here— do not have the meaning of quasiparticle states, Eq. (2) is *rigorous*, in the TDDFT linear-response framework [8]; the only approximation we have made in the evaluation of $\chi^{(s)}$ is the LDA. We have performed additional calculations within the random-phase approximation (RPA), in which *all* exchange-correlation effects are left out, including those in the band structure —which now corresponds to the Hartree approximation. Our RPA results are given on the left panel of Fig. 4, together with the full LDA results. The key physical change is that the flat Hartree *d*-bands are shifted upwards in relation to the LDA *d*-bands of Fig. 3; the former lie almost entirely *above* the shaded strip in Fig. 4. As a result, the plasmon linewidth dispersion curve acquires a much smaller slope, which differs significantly from experiment. This finding places our result of Fig. 2 in an even more interesting perspective, since it means that the correlations contained in the one-particle-like Kohn-Sham states of *d*-symmetry play a non-trivial role —via $\chi^{(s)}$— in the explanation of the plasmon damping mechanism. These correlations are usually not dealt with explicitly in many-body models of correlated electrons.

Finally, we note that K has a polarizable core, which impacts the damping process via the second factor in Eq. (3). The “local-orbital” extension of the LAPW method [12] allows us to treat the contribution to $\chi^{(s)}$ from higher-lying core states, mainly derived from the $3p^6$ atomic states (whose threshold is at 16 eV), *on the same footing* with that from the valence states [25]. The circles (squares) on the right panel of Fig. 4 are the result obtained in the presence (absence) of this contribution to $\chi^{(s)}$. It is apparent that the absolute value of the linewidth is affected markedly by the availability of the core excitations. In fact, their effect on *the slope* of the linewidth dispersion curve is even larger than that of the many-particle correlations [23]. Of course, this effect was included in our central result of Fig. 2.

In summary, we have explained the "anomalous" dispersion of the linewidth of the K plasmon via *ab initio* TDDFT-based calculations. The key mechanism was shown to be decay into particle-hole pairs involving empty states of *d*-symmetry. Our results highlight the role of the exchange-correlation effects built into the LDA single-particle band structure —these effects become important because the key final-state bands are flat. It is the phase-space complexity associated with these bands which renders simple models inapplicable. *Ab initio* paradigms should also play an important role in the study of excitations in materials with complicated band structures near the Fermi surface [26].

We thank Christian Halloy and the Joint Institute for Computational Science for training and technical support with the use of the University of Tennessee IBM SP2 computer. Thanks are also due to Wolf-Dieter Schöne for help with the Padé approximant technique, and to Ben Larson and Ward Plummer for critical comments on the manuscript. This work was supported by NSF Grant No. DMR-9634502 and the National Energy Research Supercomputer Center. ORNL is managed by Lockheed Martin Energy Research Corp. for the Division of Materials Sciences, U.S. DOE under contract DE-AC05-96OR2464.

## FIGURE CAPTIONS

Fig. 1. Loss function for K for $\vec{q} = (1,1,0)(2\pi/(16a_0))$, where $a_0 = 5.23\,\text{Å}$ is the lattice constant. Thick solid line: $\operatorname{Im}\chi$ obtained from Eqs. (1) and (2) for imaginary $\omega$'s, followed by analytic continuation to a distance $\delta$ above the real-$\omega$ axis [15]. Thin solid line: $\operatorname{Im}\chi$ obtained directly for real $\omega$'s, using a broadening parameter $\eta$ in Eq. (2). Top (bottom) panel corresponds to $\delta = \eta = 1\,\text{meV}$ ($10\,\text{meV}$).

Fig. 2. Plasmon linewidth dispersion for K. Comparison of our theoretical results with the EELS data of Ref. [2] (diamonds), and the theoretical results of Ref. [6] (solid line). Theory is for (1,1,0) propagation; the EELS data are for polycrystalline K.

Fig. 3. Left panel: Plasmon linewidth dispersion obtained upon keeping 3 (triangles) and 6 (squares) valence bands in $\chi^{(s)}$ [24]. Inset: LDA band structure of K; the arrow indicates the value of $\omega_p(0)$ (see text and [22]). Right panel: Calculated DOS for K —total DOS and contributions from states of *s*,*p*, and *d*- symmetry [11]; the zero of energy is the Fermi level.

Fig. 4. Left panel: Plasmon linewidth dispersion obtained on the basis of RPA (diamonds) and LDA (circles) single-particle response $\chi^{(s)}$ and respective band structures. The inset shows Hartree bands used in the RPA case; note the impact of exchange-correlation effects on the location of the *d*-bands (cf. Fig. 3). Right panel: Calculated plasmon linewidth dispersion in the presence (circles) or absence (squares) of the core contribution to the Kohn-Sham response $\chi^{(s)}$. See text.

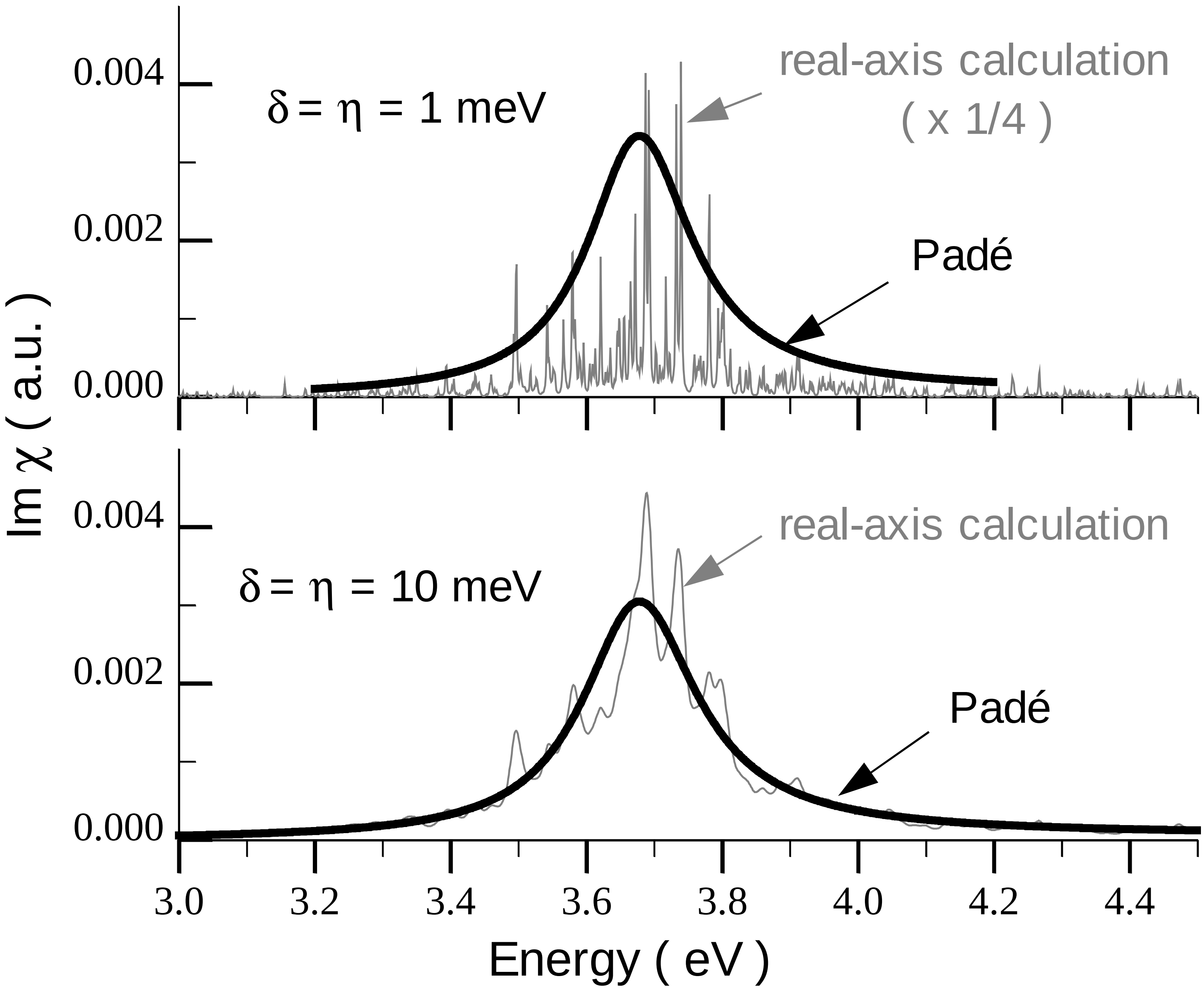

Fig. 1. Ku & Eguiluz

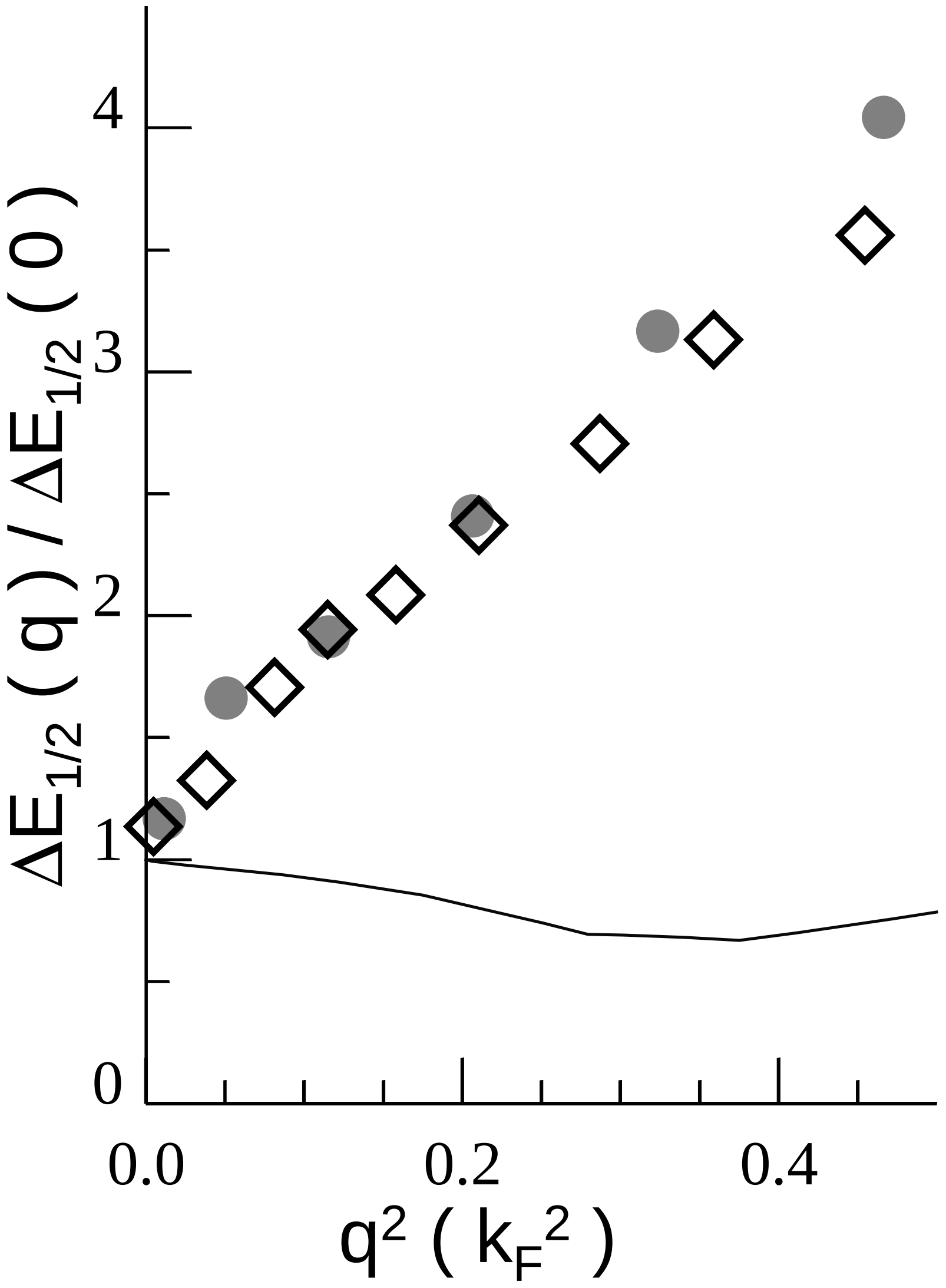

Fig. 2. Ku & Eguiluz

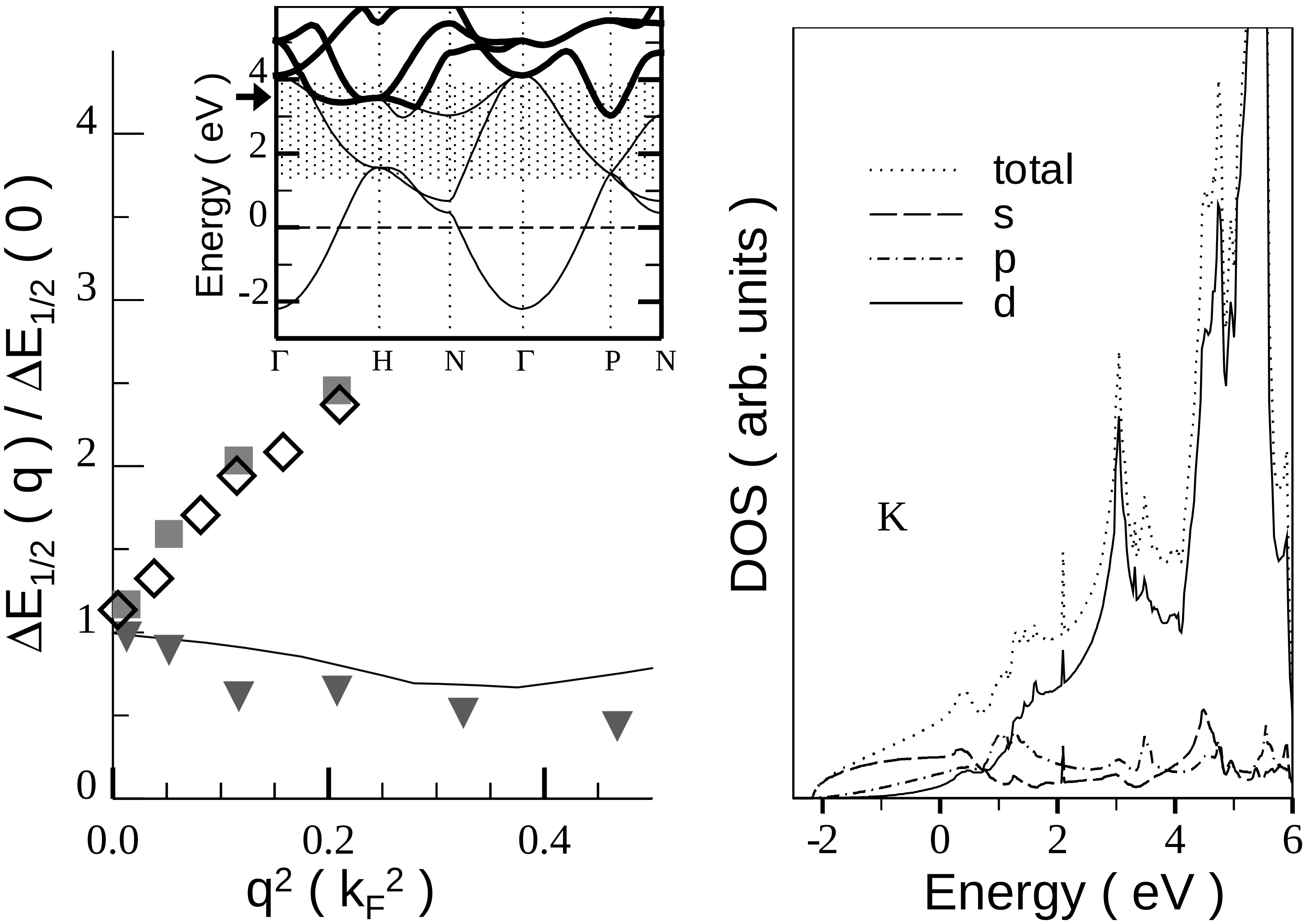

Fig. 3. Ku & Eguiluz

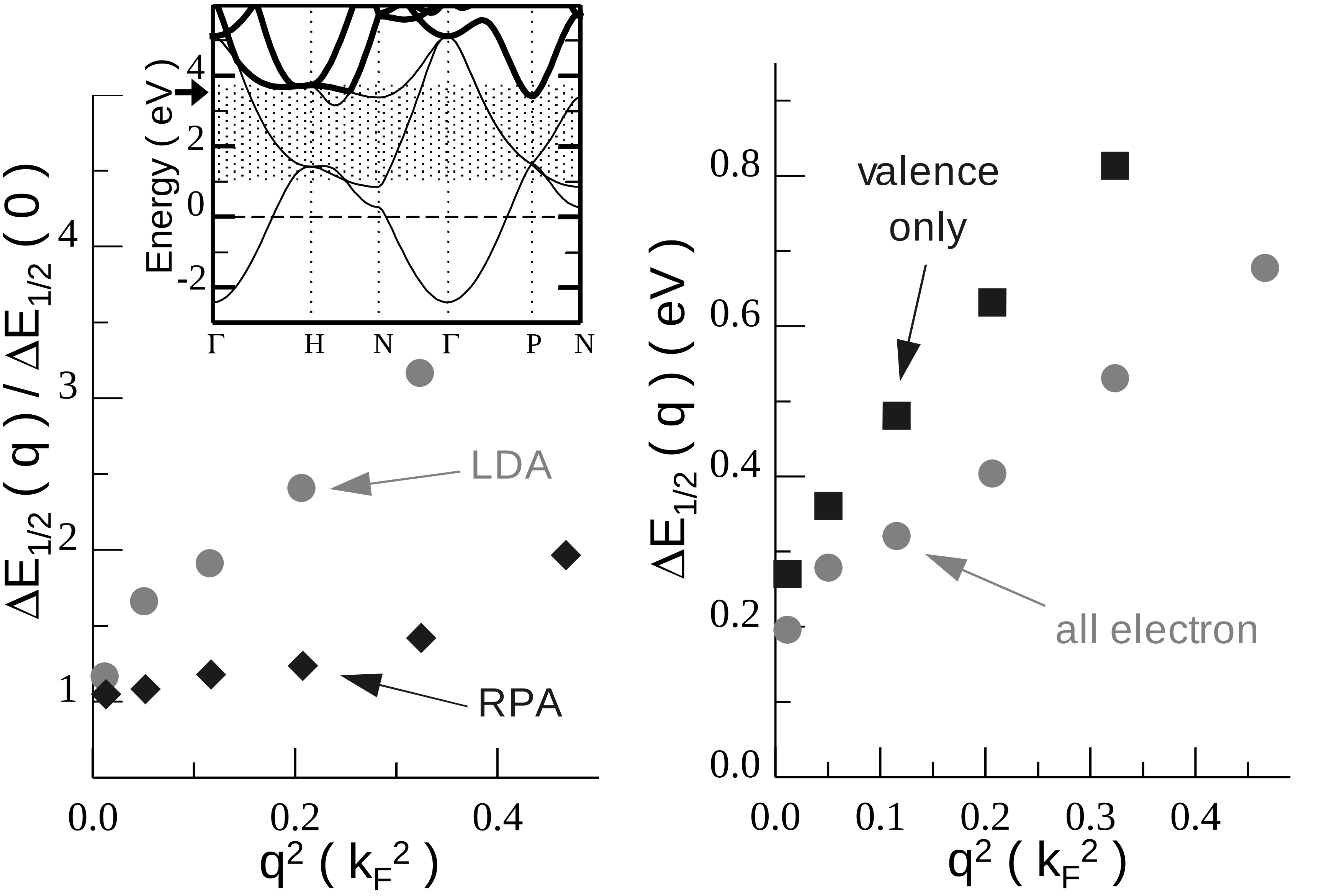

Fig. 4. Ku & Eguiluz